\documentclass{iopjournal}

\usepackage{amsfonts,amsmath,mathtools,amssymb}
\usepackage[bb=boondox]{mathalfa}

\usepackage{graphicx}
\usepackage{tikz}

\newcommand\bb[1]{\mbox{\boldmath{$#1$}}}
\newcommand\grad{\bb{\nabla}}
\newcommand\bcdot{\,\bb{\cdot}\,}

\newcommand\btimes{\,\bb{\times}\,}

\newcommand{\rmd}{\textrm{d}}
\newcommand{\rms}{\textrm{s}}

\newcommand{\vecOmz}{\bb{\Omega}_0} 
\newcommand{\Omz}{\Omega_0}

\newcommand{\hatvece}{\hat{\bb{e}}} 
\newcommand{\vecv}{\bb{v}} 
\newcommand{\vecV}{\bb{V}} 
\newcommand{\vecE}{\bb{E}} 
\newcommand{\vecB}{\bb{B}} 
\newcommand{\veceps}{\bb{\epsilon}} 
\newcommand{\vecnu}{\bb{\nu}} 
\newcommand{\vectau}{\bb{\tau}} 
\newcommand{\tilveceps}{\widetilde{\bb{\epsilon}}} 
\newcommand{\tileps}{\widetilde{\epsilon}}
\newcommand{\tilvectau}{\widetilde{\bb{\tau}}}
\newcommand{\tiltau}{\widetilde{\tau}}
\newcommand{\bgam}{\bar{\gamma}}
\newcommand{\vecJ}{\bb{J}} 
\newcommand{\vecu}{\bb{u}} 
\newcommand{\vecx}{\bb{x}} 
\newcommand{\vecF}{\bb{F}} 
\newcommand{\pd}{\partial}

\newcommand{\be}{\begin{equation}}
\newcommand{\ee}{\end{equation}}

\DeclareMathAlphabet\mathbfcal{OMS}{cmsy}{b}{n}

\begin{document}

\articletype{Paper} 

\title{Particle-in-Cell Methods for Simulations of Sheared, Expanding, or Escaping Astrophysical Plasma}

\author{Fabio Bacchini$^{1,2,*}$\orcid{0000-0002-7526-8154}, Evgeny A. Gorbunov$^3$\orcid{0000-0001-9073-8591}, Maximilien P\'eters de Bonhome$^{1,2}$\orcid{0000-0002-7493-6181}, Paul Els$^{1}$\orcid{0000-0003-2141-7705}, Konstantinos-Xanthos Argyropoulos$^1$, Minh Nhat Ly$^4$\orcid{0009-0009-6218-5446}, and Daniel Gro\v{s}elj$^{1}$\orcid{0000-0000-0000-0000}}

\affil{$^1$Centre for mathematical Plasma Astrophysics, Department of Mathematics, KU Leuven, Leuven, Belgium}

\affil{$^2$Royal Belgian Institute for Space Aeronomy, Solar-Terrestrial Centre of Excellence, Brussels, Belgium}

\affil{$^3$Department of Physics, University of Maryland, College Park, 
USA}

\affil{$^4$Institute of Laser Engineering, The University of Osaka,
Osaka,
Japan}

\email{*\href{mailto:fabio.bacchini@kuleuven.be}{fabio.bacchini@kuleuven.be}}


\begin{abstract}
Particle-in-Cell (PIC) methods have achieved widespread recognition as simple and flexible approaches to model collisionless plasma physics in fully kinetic simulations of astrophysical environments. However, in many situations the standard PIC algorithm must be extended to include macroscopic effects in microscale simulations. For plasmas subjected to shearing or expansion, shearing-box and expanding-box methods can be incorporated into PIC to account for these global effects. For plasmas subjected to local acceleration in confined regions of space, a leaky-box method can allow closed-box PIC simulations to account for particle escape from the accelerator region. In this work, we review and improve methods to include shearing, expansion, and escape in PIC simulations. We provide the numerical details of how Maxwell's equations and the particle equations of motion are solved in each case, and introduce generalized Boris-like particle pushers to solve the momentum equation in the presence of extra forces. This work is intended to serve as a comprehensive reference for the implementation of shearing-box, expanding-box, and leaky-box algorithms in PIC.
\end{abstract}

\section{Introduction}

Fully kinetic models of astrophysical plasmas have gained widespread recognition for their capability of reproducing collisionless plasma dynamics. 
Particle-in-Cell (PIC) methods \cite{Dawson1983,Birdsall2005}, in particular, are the most commonly employed numerical tool to conduct fully kinetic simulations. The simplicity and reliability of PIC methods allow for studies of a wide variety of (astrophysical) plasma phenomena, both in the nonrelativistic and relativistic regimes \cite{lapenta2012,nishikawa2021}.

Standard PIC methods include a Maxwell solver to advance the electric and magnetic fields $\vecE$, $\vecB$ in time as well as a particle pusher to propagate particles in position--velocity phase space $(\vecx,\vecu)$ according to the charged-particle equations of motion. Moments of the velocity distribution function are collected from the particles onto a numerical grid, to construct the sources of Maxwell's equations; the electromagnetic fields, in turn, are interpolated from the grid to the particles to determine the particle motion through the Lorentz force. 

For plasmas embedded within realistic astrophysical environments, however, the relatively simple algorithm described above may not suffice. Particularly in 
local boxes, there is a need to couple local plasma evolution to the global-scale configuration of the system and its surrounding medium.
This includes, for instance, the effects of background shear in local accretion-disk simulations; the modeling of local plasma 
expansion in magnetized winds launched from a central source; or the inclusion of diffusive cosmic-ray escape in local boxes.
The PIC method needs modifications to be able to model kinetic plasmas in such scenarios, which are the subject of ongoing investigations. 

For plasmas that experience shearing forces, the \textit{shearing-box method} has been used in kinetic simulations primarily to model plasma turbulence in accretion disks. The general approach is to simulate an initially thermal plasma threaded by a weak magnetic field and subjected to background shear mimicking the differential rotation of the disk. This allows the development of the magnetorotational instability (MRI) \cite{balbushawley1991}, which nonlinearly develops into sustained turbulence. The shearing, and potentially Coriolis and gravitational forces, are incorporated into Maxwell's equations and into the particle equations of motion to augment standard PIC algorithms. Initial works using the shearing box focused on 2D systems with or without ``shearing coordinates'' \cite{riquelme2012,hoshino2013,inchingolo2018}. Pioneering 3D simulations were also conducted with relatively small system sizes \cite{hoshino2015}. However, only hybrid simulations (discarding electron physics) employed large enough 3D system sizes that could retrieve the expected sustained-turbulence nonlinear state \cite{kunz2016}. In fully kinetic runs, this expected behavior was first reported by us \cite{bacchini2022mri,bacchini2024prl} for the pair-plasma case and later extended to incorporate radiation and ion--electron physics \cite{gorbunov2025apjl}. The stratified-density case was also recently simulated for the first time with fully kinetic methods in 2D and 3D \cite{sandoval2024}.

For plasmas that are subjected to expansion or compression, the \textit{expanding-box method} has been designed to enable local kinetic simulations of e.g.\ expanding solar-wind plasma or compressing plasma in accretion disks or the intracluster medium. In PIC runs, expansion/contraction can be included by modifying Maxwell's equations to take length dilation/contraction into account in the curl terms, and by adding extra forces on the particle equations of motion. 
Most previous works employing the expanding box have focused on hybrid simulations (treating electrons as a massless fluid) \cite{hellinger2019,bott2021}; fully kinetic runs, including electron physics, have rarely been explored, with notable exceptions for the expanding solar wind \cite{innocenti2019method,innocenti2019,micera2021} and for compressing plasma in accretion disks and galaxy clusters \cite{sironinarayan2015, tran2023}.

In plasmas where particles are subjected to efficient acceleration, e.g.\ in the surroundings of compact astrophysical objects, a typical approach is to consider a PIC model of a local turbulent box. In such models, turbulence is usually driven by some external force, and the plasma continuously heats up due to the turbulent cascade. Accelerated particles are expected to leave the turbulent region by diffusing out of the source, carrying away energy. To model these dynamics and attain a steady state without secular, unbound energy pile-up, a \textit{leaky-box approach} can be used \cite{gorbunov2025prl,groselj2026}. In this case, particles can leave the simulation box according to some prescription, potentially reaching a balance between energy injection and escape.

In this paper, we detail some of our recent advances in modeling plasmas in realistic astrophysical environments. In particular, we review and improve upon PIC methods to model shearing (Section~\ref{sec:shear}), expansion (Section~\ref{sec:expand}), and particle escape (Section~\ref{sec:leak}) in local, fully kinetic plasma simulations. We provide numerical details for each method and present example applications. We finally summarize our results and discuss future perspectives (Section~\ref{sec:conclusion}).

\section{Shearing Box for PIC}
\label{sec:shear}

\subsection{Summary of the Kinetic Shearing Box with Orbital Advection}
The \textit{kinetic shearing box with orbital advection} (KSB-OA) was introduced recently \cite{bacchini2022mri} as an alternative to preexisting methods \cite{riquelme2012,hoshino2013,hoshino2015}. The KSB-OA strategy is applicable to 3D shearing-box simulations of collisionless plasmas and is based on the orbital-advection approach applied in magnetohydrodynamic shearing-box calculations \cite{gresselziegler2007,stonegardiner2010}.

In the KSB-OA formulation, the simulation frame corresponds to the corotating frame, i.e.\ the point of view of an observer orbiting around a central object. In this frame, we employ coordinates $(x,y,z)$ where $r+x$ gives the local radial position around $r$, $y$ represents the toroidal direction, and $z$ is the vertical direction. Furthermore, it is assumed that:
\begin{itemize}
    \item The orbital (rotational) velocity $\vecv_0=\bb{\Omega}_0\btimes\bb{r}$, where $\bb{\Omega}_0=(0,0,\Omz)$ is the rotational frequency, is such that $v_0\ll c$ at the radius $r_0$ where the observer lies.
    \item The rotational profile $\Omz(r)$ is such that adjacent plasma parcels rotate differentially, imposing a shear on the plasma.
    \item The physical system under consideration has linear dimension $\ll r_0$, allowing us to approximate the local shearing velocity as $\vecv_\rms=-s\Omz x\hatvece_y$, with $s=3/2$ for a Keplerian profile.
\end{itemize}
Because $v_0\ll c$, the electromagnetic-field transformation to the corotating frame is essentially ignored and Maxwell's equations retain their standard form. The particle equations of motion acquire contributions from the Coriolis, centrifugal, and radial gravitational forces\footnote{Shearing-box simulations that include vertical density stratification also add vertical gravitational forces \cite{sandoval2024}, which we ignore.}. Under the assumptions above, we construct the KSB-OA equations by further transforming the electric field and particle momenta to the comoving (shearing) frame\footnote{This approach is also inspired by nonrelativistic hybrid-PIC approaches \cite{kunz2014code}.} with velocity $\vecv_\rms$:
\be
\vecE' = \vecE + \frac{\vecv_\rms}{c}\btimes\vecB,
\ee
\be
\vecu' = \vecu-\gamma'\vecv_\rms,
\ee
where $\vecE'$ and $\vecu'$ are the comoving electric field and spatial part of the particle four-velocity, and $\gamma'=\sqrt{1+(u'/c)^2}$. For both transformations, we have assumed that $v_\rms\ll c$ and therefore $\gamma'\simeq\gamma=\sqrt{1+(u/c)^2}$. Particle positions and magnetic fields are not transformed. The current density transforms accordingly,
\be
\vecJ' = \vecJ-\rho\vecv_\rms,
\ee
with $\rho\equiv q\int f\rmd^3\vecu$ the charge density and $\vecJ\equiv q\int(\vecu/\gamma) f\rmd^3\vecu$ (and equivalently for $\vecJ'$) for a particle species with charge $q$ and distributed according to $f$. The resulting KSB-OA equations for the electromagnetic fields are
\begin{equation}
 \pd_t\vecB = -c\grad\btimes\vecE' - (\vecv_\rms\bcdot\grad)\vecB - s\Omz B_x\hatvece_y,
 \label{eq:OAB}
\end{equation}
\begin{equation}
 \pd_t\vecE' = c\grad\btimes\vecB - 4\pi\vecJ' - \vecv_\rms\btimes(\grad\btimes\vecE') - \vecv_\rms \grad\bcdot\vecE',
\label{eq:OAE}
\end{equation}
and for the particles we have
\be
\frac{\rmd\vecx}{\rmd t} = \frac{\vecu'}{\gamma'} + \vecv_\rms,
\label{eq:OAx}
\ee
\be
\frac{\rmd\vecu'}{\rmd t} = \frac{q}{m}\left(\vecE'+\frac{\vecu'}{c\gamma'}\btimes\vecB\right) + 2\vecu'\btimes\vecOmz + s\Omz u_x'\hatvece_y - \frac{q}{m}\left(\frac{\vecu'}{c\gamma'}\bcdot\vecE'\right)\frac{\vecv_\rms}{c}.
\label{eq:OAu}
\ee
In deriving these equations, we have adopted further assumptions: i) we ignored higher-order terms in $\Omz$; ii) we have assumed a nonrelativistic velocity transformation such that particles move with an effective velocity $\vecu'/\gamma'+\vecv_\rms$. The latter assumption may result in superluminal motion for relativistic particles, and therefore needs special treatment, which we have discussed in the past \cite{bacchini2022mri}.

Employing $\vecE'$, $\vecB$, $\vecu'$, and $\vecx$ as evolving variables has several advantages for numerical simulations. First, the boundary conditions for grid-defined quantities are
\begin{equation}
 \vecE'(x_\mathrm{i},y,z) = \vecE'(x_\mathrm{o},y-\Delta y_\mathrm{s},z),
\end{equation}
\begin{equation}
 \vecB(x_\mathrm{i},y,z) = \vecB(x_\mathrm{o},y-\Delta y_\mathrm{s},z),
\end{equation}
\begin{equation}
 \vecJ'(x_\textrm{i},y,z) = \vecJ'(x_\textrm{o},y-\Delta y_\mathrm{s},z),
\end{equation}
where $x_\mathrm{i}$ and $x_\mathrm{o}$ are the inner and outer edges of the domain along $x$, and $\Delta y_\rms=s\Omz(x_\mathrm{o}-x_\mathrm{i})t$ is the displacement along $y$ due to the background shearing at time $t$. Therefore, for standard simulations with shearing-periodic boundaries along $x$ and standard periodic boundaries along $y$ and $z$, $\vecE'$, $\vecB$, and $\vecJ'$ simply need to be properly matched across the $x$-boundary by taking into account the displacement $\Delta y_\rms$. In contrast, $\vecE$ and $\vecJ$ would additionally require a modification determined by the local $\vecB$ and $\rho$, as can be seen from the transformations discussed above. Second, by using $\vecu'$, we can directly compute $\vecJ'$, as we also discuss below.

\subsection{Numerical Solution of the KSB-OA Equations}
In comparison with standard PIC, solving the KSB-OA equations involves two main complications: Maxwell's equations become implicit, and the particle equations of motion contain more terms.

Maxwell's equations \eqref{eq:OAB}--\eqref{eq:OAE} for $\vecE'$ and $\vecB$ cannot be easily treated with an explicit leapfrog scheme as in standard PIC, because of the extra terms on the right-hand side. Therefore, our Maxwell solver employs a standard implicit midpoint method in time for both equations, and a central-difference scheme in space. An important exception is made for advection terms: it can indeed be noted that eqs.~\eqref{eq:OAB}--\eqref{eq:OAE} can be written in the form
\begin{equation}
 \pd_t\vecB = - (\vecv_\rms\bcdot\grad)\vecB + \bb{F}_{\boldsymbol{B}},
\end{equation}
\begin{equation}
 \pd_t\vecE' = \pm (\vecv_\rms\bcdot\grad)\vecE' + \bb{F}_{\boldsymbol{E}'},
\end{equation}
where ``$\pm$'' in the equation for $\vecE'$ is a ``$-$'' for the $y$-component and a ``$+$'' for the others, and $\vecF_{\boldsymbol{B},\boldsymbol{E}'}$ contains the rest of the right-hand-side terms. Given the form of $\vecv_\rms$, the advection terms on the right-hand side represent a transport of the fields along $y$ due to the background shear. These terms need special care because of potential numerical instabilities; we have verified that the most robust approach involves spatial upwinding in the local direction of $\vecv_\rms$, which we therefore apply only to the advection terms. The resulting implicit set of equations is solved with a fixed-point iterative scheme, which typically requires only a few iterations to converge below acceptable error thresholds.

Particle position and momenta are leapfrog-discretized in time. To solve the momentum equation~\eqref{eq:OAu}, we discretize it in a Boris-like fashion \cite{boris1970} between two consecutive time steps $n-1/2,n+1/2$ and recast it as 
\begin{equation}
  \vecu^+ - \vecu^+\btimes\tilvectau - \alpha_1 u^+_x\hatvece_y - \alpha_2 (\vecu^+\bcdot\tilveceps)\hatvece_y = 
  \vecu^- + \vecu^-\btimes\tilvectau + \alpha_1 u^-_x\hatvece_y + \alpha_2 (\vecu^-\bcdot\tilveceps)\hatvece_y,
  \label{eq:OAu_disc}
\end{equation}
where we have defined 
$$
\begin{aligned}
\vecu^\pm & =\vecu'^{,n\pm1/2}\mp\veceps, & \quad & \\
\veceps & = q\Delta t\vecE'/(2m), & \quad \tilveceps & = \veceps/(c\bgam), \\
\vectau & = q\Delta t\vecB/(2mc), & \quad \tilvectau & = \vectau/\bgam+\Delta t\vecOmz, \\
\alpha_1 & = s\Omz\Delta t/2, &
\quad \alpha_2 &= s\Omz x^n/c,
\end{aligned}
$$
and $\vecu^+$ becomes the unknown to solve for. Assuming that the Lorentz factor at the midpoint in time $\bgam$ is known (see below), eq.~\eqref{eq:OAu_disc} has an explicit Boris-like solution,
\be
\vecu^+ = \mathcal{B}(\vecu^-) + \mathcal{R}(S\vecu^-) + \frac{1}{(1-g)}\mathcal{R}(S(\mathcal{B}(\vecu^-) + \mathcal{R}(S\vecu^-))),
\ee
where the first term is the standard Boris operator, which reads
\be
\mathcal{B}(\vecnu) = \vecnu + 2\frac{\vecnu \btimes\tilvectau}{1+\tiltau^2} + 2\frac{(\vecnu\btimes\tilvectau)\btimes\tilvectau}{1+\tiltau^2},
\ee
for a generic vector $\vecnu$; the second term is a ``half-Boris'' rotation operator,
\be
\mathcal{R}(\vecnu) = \vecnu + \frac{\vecnu \btimes\tilvectau}{1+\tiltau^2} + \frac{(\vecnu\btimes\tilvectau)\btimes\tilvectau}{1+\tiltau^2},
\ee
and with these definitions, $g=S_{2}\bcdot\mathcal{R}(\hatvece_y)$. In all the above, $S$ is a rank-1 matrix encoding the $\alpha$-terms in eq.~\eqref{eq:OAu_disc},
\begin{equation}
    S = 
    \begin{pmatrix}
        0 & 0 & 0 \\
        \alpha_1 + \alpha_2\tileps_x & \alpha_2\tileps_y & \alpha_2\tileps_z \\
        0 & 0 & 0
    \end{pmatrix}.
\end{equation}
After $\vecu^+$ has been calculated, the particle velocity at the next time step is simply
\begin{equation}
\vecu'^{,n+1/2} = \vecu^+ + \veceps.
\end{equation}
With this sequence of operations, we notice that when $\alpha_1=\alpha_2=0$ (no shearing), $S=\mathbb{0}$ and we retrieve the pure Boris algorithm.

The definition of $\bgam$ to be employed for the particle push above is not unique \cite{ripperda2018a}. For simplicity, we take the Boris-like expression
\begin{equation}
\bgam = \sqrt{1 + \left[(u^-)^2 + 2\alpha_1 u^-_x u^-_y + 2\alpha_2(\vecu^-\bcdot\veceps) u^-_y/\bgam\right]/c^2},
\end{equation}
with which we again retrieve the standard Boris pusher when $\alpha_1=\alpha_2=0$. The resulting cubic equation for $\bgam$ can be solved with any preferred algorithm for the (only) real root $\bgam\ge1$. A good approximation for this root (in the limit of small $\Omz$) is
\be
\bgam\approx\gamma^- + \frac{\alpha_1u_x^-u_y^-\gamma^-+\alpha_2(\vecu^-\bcdot\veceps)u_y^-}{(c\gamma^-)^2},
\ee
where $\gamma^-=\sqrt{1+(u^-/c)^2}$.

With the new particle velocity, we push particle positions in two steps. The position is first updated with the motion with respect to the background, i.e.\
\begin{equation}
\vecx' = \vecx^n + \Delta t \frac{\vecu'^{,n+1/2}}{\gamma'^{,n+1/2}},
\end{equation}
with which the comoving current $\vecJ'$ can be collected from the displacement $\vecx'-\vecx^n$ \cite{villasenorbuneman1992}. Then, the shearing-velocity contribution is added,
\begin{equation}
\vecx^{n+1} = \vecx' - s\Omz \frac{x'+x^n}{2} \Delta t\hatvece_y,
\end{equation}
completing the position-update step. Note that collecting the current as mentioned above provides $\vecJ'(\vecx')$; computing $\vecJ'(\vecx)$, i.e.\ the comoving current in the lab-frame coordinates where the computational grid is defined, further requires advecting $\vecJ'(\vecx')$ along $y$ by a distance $v_{\rms,y}\Delta t/2$.

In summary, the KSB-OA algorithm is composed of the following steps:
\begin{enumerate}
    \item Interpolate electromagnetic fields at the particle position $\vecx^n$ and evolve the particle velocity $\vecu'^{,n-1/2}\to\vecu'^{,n+1/2}$ using the Boris-like algorithm above.
    \item Evolve the particle position partially, $\vecx^n\to\vecx'$, with respect to the background shear.
    \item Collect the current $\vecJ'$ from the displacement $\vecx'-\vecx^n$ and advect it along $y$ to account for the background shear transport.
    \item Fully update particle positions, $\vecx'\to\vecx^{n+1}$, by adding the background shear motion.
    \item Solve Maxwell's equations with a fixed-point iteration.
\end{enumerate}


\subsection{Applications of the PIC Shearing Box}

\begin{figure*}[t]
    \centering 
    \includegraphics[width=1\textwidth]{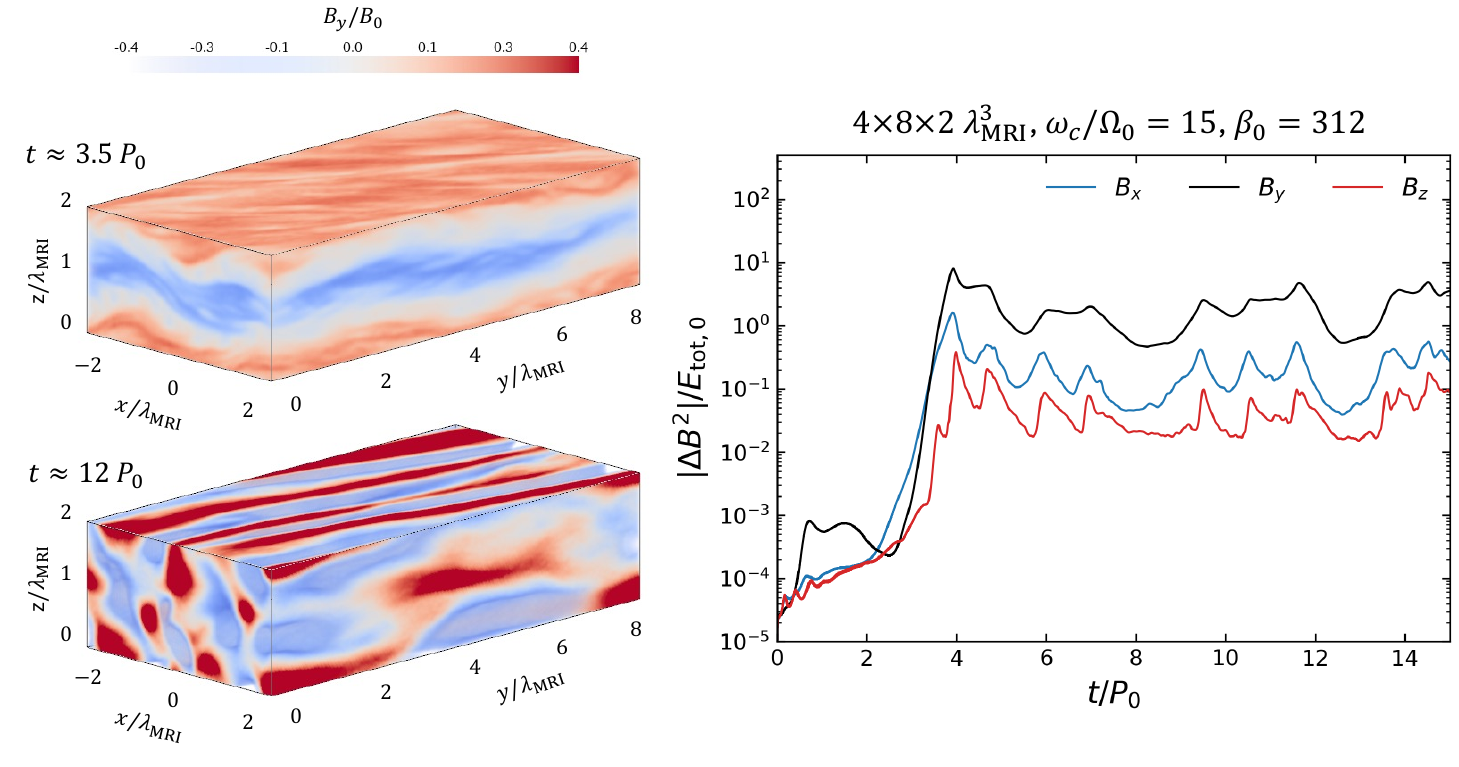}
    \caption{Example PIC simulation of the collisionless pair-plasma MRI, using the KSB-OA method. Left: spatial distribution of the toroidal ($B_y$) magnetic field during the linear MRI stage ($t \approx 3.5P_0$) and during the turbulent, nonlinear quasi-steady state ($t \approx 12P_0$). Right:  evolution of the volume-averaged change in magnetic energy for each component of $\vecB$.}
    \label{fig:shearingbox}
\end{figure*}

The KSB-OA method described above can be applied to study the evolution of the collisionless MRI. Fig.~\ref{fig:shearingbox} presents a 3D fully kinetic shearing-box example simulation of magnetorotational turbulence in a collisionless pair plasma.
The simulation domain size is based on the most-unstable MRI wavelength $\lambda_{\rm MRI} \equiv 2\pi v_{\rm A,0} /\Omega_0$ as $L_x \times L_y \times L_z = 2 \times 4 \times 8 \lambda_{\rm MRI}^3$, where $v_{\rm A,0}$ is the initial Alfv\'en speed $v_{\rm A,0}/c \equiv B_0/\sqrt{4\pi n_0 m_e c^2}$.
The simulation is initialized with a purely vertical magnetic field $\bb{B}_0=(0,0,B_0)$ and a uniform density $n_0/2$ for each species. Each plasma species is initialized from a Maxwellian distribution with temperature $\theta_0 \equiv k_\mathrm{B}T_0/(m_ec^2)=1/128$, and the initial Alfv\'en speed is set to $v_{\rm A,0}/c \simeq 0.007$. This gives an initial (total) plasma-$\beta \equiv n_0k_\mathrm{B}T_0/(B_0^2/(8 \pi)) \approx 312$.

Fig.~\ref{fig:shearingbox} (left) shows the spatial distribution of $B_y$ at $t\approx 3.5P_0$ (top) and $t\approx 12P_0$, with $P_0\equiv2\pi/\Omega_0$ the rotational period. The MRI evolves through the phases expected from fluid calculations \cite{balbushawley1991}, first creating channel flows and then entering a turbulent nonlinear stage. Fig.~\ref{fig:shearingbox} (right) shows the evolution in time of the volume-averaged magnetic energy in each component of $\vecB$. The shearing conditions applied to the $x$-boundary amplify the initial magnetic field exponentially in the linear  MRI stage, around $t\approx 2\mbox{--}4P_0$; for $t \gtrsim 4P_0$, the MRI saturates into a quasi-steady-state turbulence with the magnetic field components fluctuating periodically around constant values. This sustained-turbulence behavior results from the relatively large box size and the 3D geometry \cite{bacchini2022mri}.

\section{Expanding Box for PIC}
\label{sec:expand}
\subsection{Summary of the Kinetic Expanding Box}

The \textit{kinetic expanding box} (KEB) has been developed by taking inspiration from similar MHD strategies \cite{squire2020} suitable to follow the evolution of a plasma parcel that expands (or contracts) over time. A number of formulations exist to treat expansion in a fully kinetic system \cite{sironinarayan2015,innocenti2019method}. Here, we broadly follow the derivation presented in ref.~\cite{sironinarayan2015}, with some modification.

In this KEB formulation, the simulation frame corresponds to the comoving frame, i.e.\ the point of view of an observer traveling with velocity $\vecV_0$, typically along the direction of a background magnetic field $\vecB_0$ (as appropriate for e.g.\ the solar wind). In this frame, we employ coordinates $(x,y,z)$ with $\vecV_0$ and $\vecB_0$ typically aligned with one of the coordinate axes. In what follows, it is assumed that:
\begin{itemize}
    \item The background velocity is such that $V_0\ll c$.
    \item In the comoving frame, the plasma expands or contracts with velocity $\bb{U}_0$ such that $V_0<U_0\ll c$.
    \item The expansion/contraction speed changes slowly over time (if at all).
\end{itemize}
Because $V_0\ll c$, the electromagnetic-field transformation to the comoving frame is essentially ignored and Maxwell’s equations retain their standard form, and the comoving frame becomes our laboratory frame. From here, we make a second transformation to the expanding/contracting frame: the coordinates transform as
\be
L = \frac{\partial \vecx}{\partial\vecx'}=
\begin{pmatrix}
    l_1 & 0 & 0 \\
    0 & l_2 & 0 \\
    0 & 0 & l_3
\end{pmatrix},
\ee
where the $l_i$ are the expansion/contraction rates along each direction, linking the lab-frame coordinates $\vecx$ to the coordinates $\vecx'$ where the expansion/contraction is effectively not seen, making the $\vecx'$ frame convenient for numerical simulations.

The lab-frame electromagnetic fields and particle momenta transform to the expanding/contracting frame as 
\be
\vecB' = \ell L^{-1} \vecB, \quad \vecE'=\ell L^{-1}\vecE,
\ee
where $\ell=\mathrm{det}(L)$, and
\be
\vecv'=L^{-1}\vecv,
\ee
where $\vecv=\vecu/\gamma$. The current density then transforms as
\be
\vecJ' = L^{-1}\vecJ,
\ee
where $\vecJ\equiv q\int(\vecu/\gamma) f\rmd^3\vecu$ in the lab-frame coordinates. With these definitions, the equations to be solved become
\be
\partial_{t'}\vecB' = -c\grad'\btimes\left(\frac{1}{\ell}L^2\vecE'\right),
\label{eq:EBB}
\ee
\be
\partial_{t'}\vecE' = c\grad'\btimes\left(\frac{1}{\ell}L^2\vecB'\right) - 4\pi\ell\vecJ',
\label{eq:EBE}
\ee
for the electromagnetic fields, and
\be
\frac{\rmd\vecx'}{\rmd t'} = L^{-1}\frac{\vecu}{\gamma},
\ee
\be
\frac{\rmd\vecu}{\rmd t'} = \frac{q}{m}\left(\vecE+\frac{\vecu}{c\gamma}\btimes\vecB\right)-L^{-1}\frac{\rmd L}{\rmd t}\vecu,
\label{eq:EBu}
\ee
for particle evolution.
Notice that we solve for $\vecx'$, the expanding/compressing-frame particle coordinates where the simulation box lives, and for $\vecu$, the lab-frame velocity. With this ansatz, collecting the current from the particles with a standard charge-conserving deposition directly provides $\ell\vecJ'$, the source term needed for Maxwell's equations. This choice is convenient for numerical simulations, as we also elaborate on below.

\subsection{Numerical Solution of the KEB Equations}
By formulating the KEB equations in terms of $\vecB$, $\vecE$, $\vecx'$, and $\vecu$, the numerical solution of the KEB system requires minimal modification with respect to a standard PIC run. The key concept is that all quantities are defined at the $\vecx'$ coordinates, but electromagnetic fields and momenta keep their values as measured in the lab frame, and can therefore be used for analysis. The coordinates $\vecx'$ do not ``feel'' the expansion, and standard boundary conditions can be applied.

Maxwell's equations can be treated with a standard Maxwell solver. At each time step, we transform $\vecE,\vecB\to\vecE',\vecB'$ and then solve eqs.~\eqref{eq:EBB}--\eqref{eq:EBE} with a standard leapfrog method by modifying $\vecE'\to (1/\ell)L^2\vecE'$ and $\vecB'\to (1/\ell)L^2\vecB'$ to be used in the curl terms on the right-hand side. The updated $\vecE',\vecB'$ fields are then transformed back to the lab-frame $\vecE,\vecB$.

To solve the momentum eq.~\eqref{eq:EBu}, we discretize it in a Boris-like fashion \cite{boris1970} between two consecutive time steps $n-1/2,n+1/2$ and recast it as 
\begin{equation}
  \vecu^+ - \Lambda\vecu^+ - \vecu^+\btimes\vectau = 
  \vecu^- + \Lambda\vecu^- + \vecu^-\btimes\vectau,
  \label{eq:EBu_disc}
\end{equation}
where we have defined
$$
\begin{aligned}
\vecu^\pm & =\vecu^{n\pm1/2}\mp\veceps, \\
\veceps & = q\Delta t\vecE/(2m), \\
\vectau & = q\Delta t\vecB/(2\bgam mc), \\ 
\Lambda & = -(\Delta t/2)L^{-1}\rmd L/\rmd t,
\end{aligned}
$$
and $\vecu^+$ is the unknown to solve for. Assuming that the Lorentz factor at the midpoint in time $\bgam$ is known (see below), eq.~\eqref{eq:EBu_disc} has an explicit Boris-like solution valid for diagonal $\Lambda$ (which holds for pure compression/expansion),
\be
\vecu^+ = \mathcal{B}^{(\Lambda)}(\vecu^-),
\ee
where we have defined a modified Boris operator $\mathcal{B}^{(\Lambda)}$ which reads
\be
\mathcal{B}^{(\Lambda)}(\vecnu) = \tilde{\Lambda}\vecnu + \frac{(\mathbb{1}+\tilde{\Lambda})\vecnu \btimes\tilvectau}{1+\tiltau^2} + \frac{((\mathbb{1}+\tilde{\Lambda})\vecnu\btimes\tilvectau)\btimes\tilvectau}{1+\tiltau^2},
\ee
for a generic vector $\vecnu$. Here, $\tilvectau = (\mathbb{1}-\Lambda)^{-1}\vectau$ (i.e.\ $\tiltau_i = \tau_i/(1-\Lambda_{ii})$) and $\tilde{\Lambda} = (\mathbb{1}-\Lambda)^{-1}(\mathbb{1}+\Lambda)$ (i.e.\ $\tilde{\Lambda}_{ii} = (1+\Lambda_{ii})/(1-\Lambda_{ii})$).  After $\vecu^+$ has been calculated, the particle velocity at the next time step is simply
\begin{equation}
\vecu^{n+1/2} = \vecu^+ + \veceps.
\end{equation}
With this sequence of operations, we notice that when $\Lambda_{ii}=0$ (no expansion/contraction), $\tilde{\Lambda}=\mathbb{1}$, and we retrieve the pure Boris algorithm.

The definition of $\bgam$ to be employed for the particle push above is not unique \cite{ripperda2018a}. Here, we take the reasonable Boris-like definition
\be
\bgam\equiv\sqrt{1+((\mathbb{1}+\Lambda)\vecu^-)^2/c^2},
\ee
with which we again retrieve the standard Boris pusher when $\Lambda=\mathbb{0}$.

With the new particle velocity, we push particle positions straightforwardly as
\be
\vecx'^{,n+1} = \vecx'^{,n} + \Delta t (L^{-1})^{n+1/2}\frac{\vecu^{n+1/2}}{\gamma^{n+1/2}},
\ee
with which the expanding/contracting-frame current density $\ell\vecJ'$ can be retrieved.

In summary, the KEB algorithm is composed of the following steps:
\begin{itemize}
\item Interpolate lab-frame electromagnetic fields $\vecB,\vecE$ at the particle position $\vecx'{^n}$ and evolve the particle velocity $\vecu^{n-1/2}\to\vecu^{n+1/2}$ using the Boris-like algorithm above.
\item Evolve the particle position $\vecx'^{,n}\to\vecx'^{,n+1}$.
\item Collect the current $\ell\vecJ'$ from the displacement $\vecx'^{,n+1}-\vecx'^{,n}$.
\item Transform $\vecB,\vecE\to\vecB',\vecE'$, solve Maxwell's equations with a standard leapfrog step, and transform the updated $\vecB',\vecE'\to\vecB,\vecE$.
\end{itemize}


\subsection{Applications of the PIC Expanding Box}

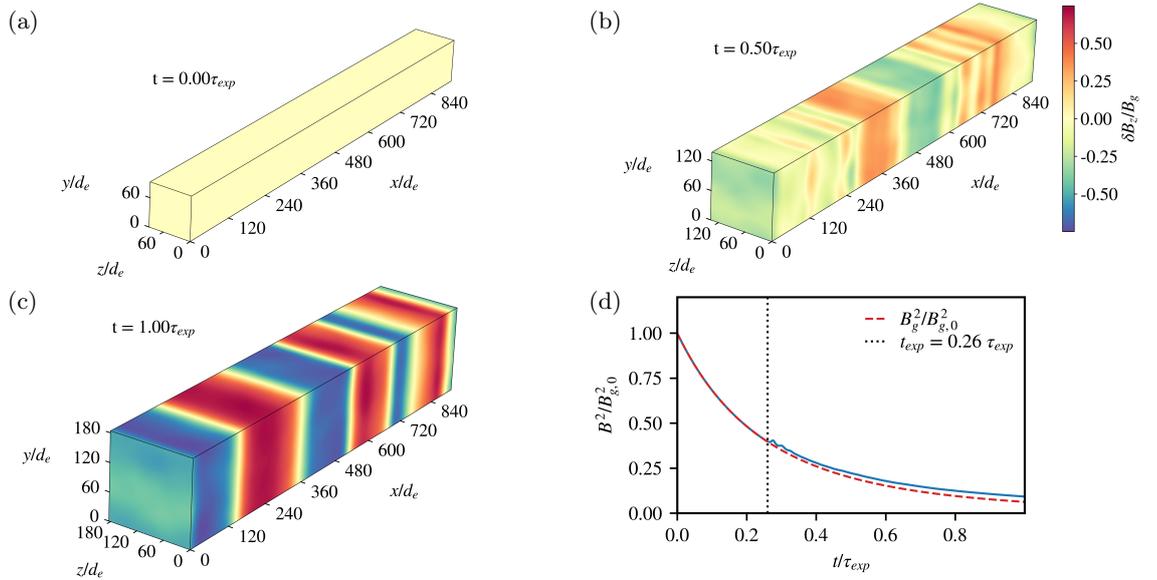
\begin{figure}[t]
\centering

\begin{minipage}[t]{0.5\linewidth}
\centering
\begin{tikzpicture}
\node[anchor=south west, inner sep=0] (img)
  {\includegraphics[width=\linewidth]{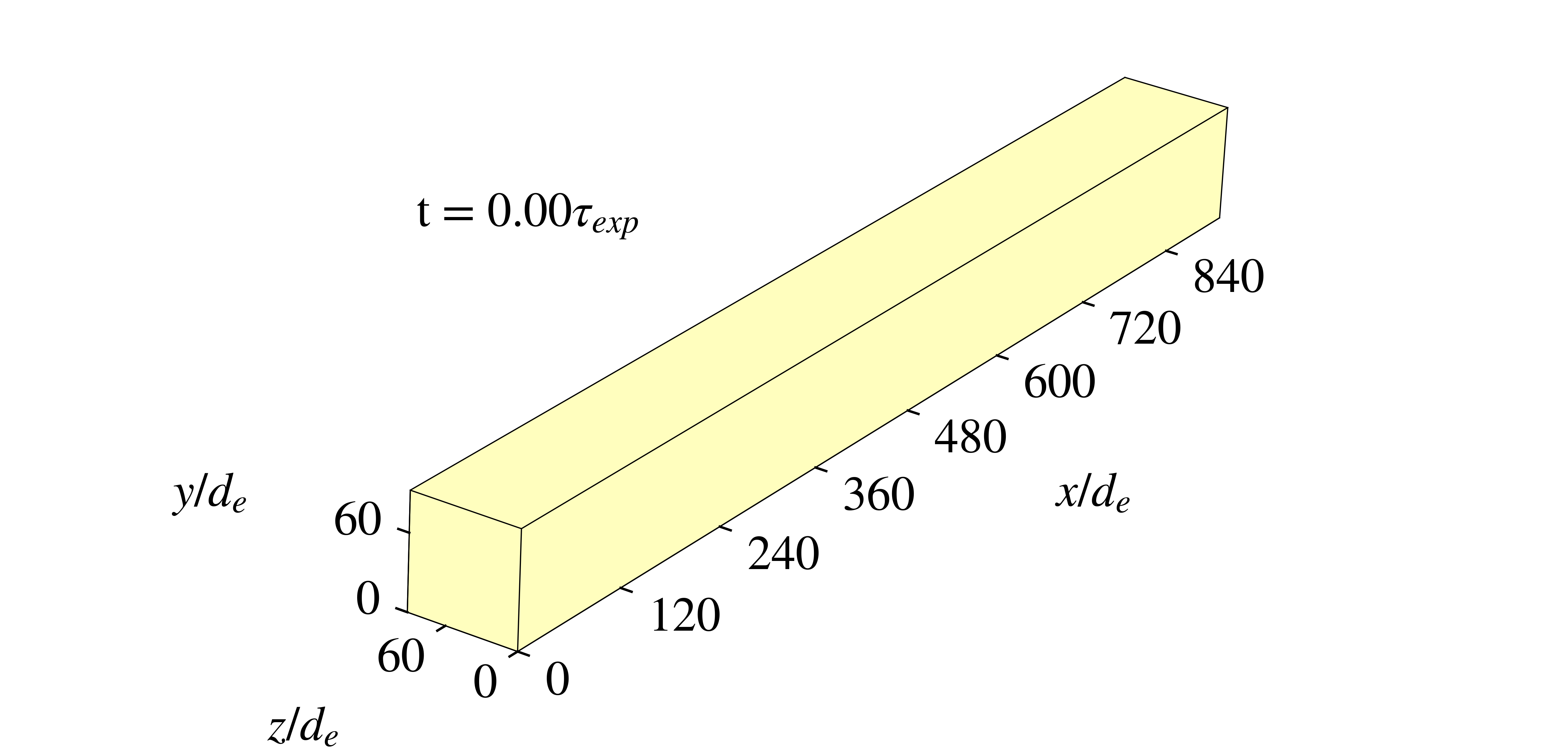}};
\node[anchor=north west] at (img.north west) {\small (a)};
\end{tikzpicture}
\end{minipage}%
\begin{minipage}[t]{0.5\linewidth}
\centering
\begin{tikzpicture}
\node[anchor=south west, inner sep=0] (img)
  {\includegraphics[width=\linewidth]{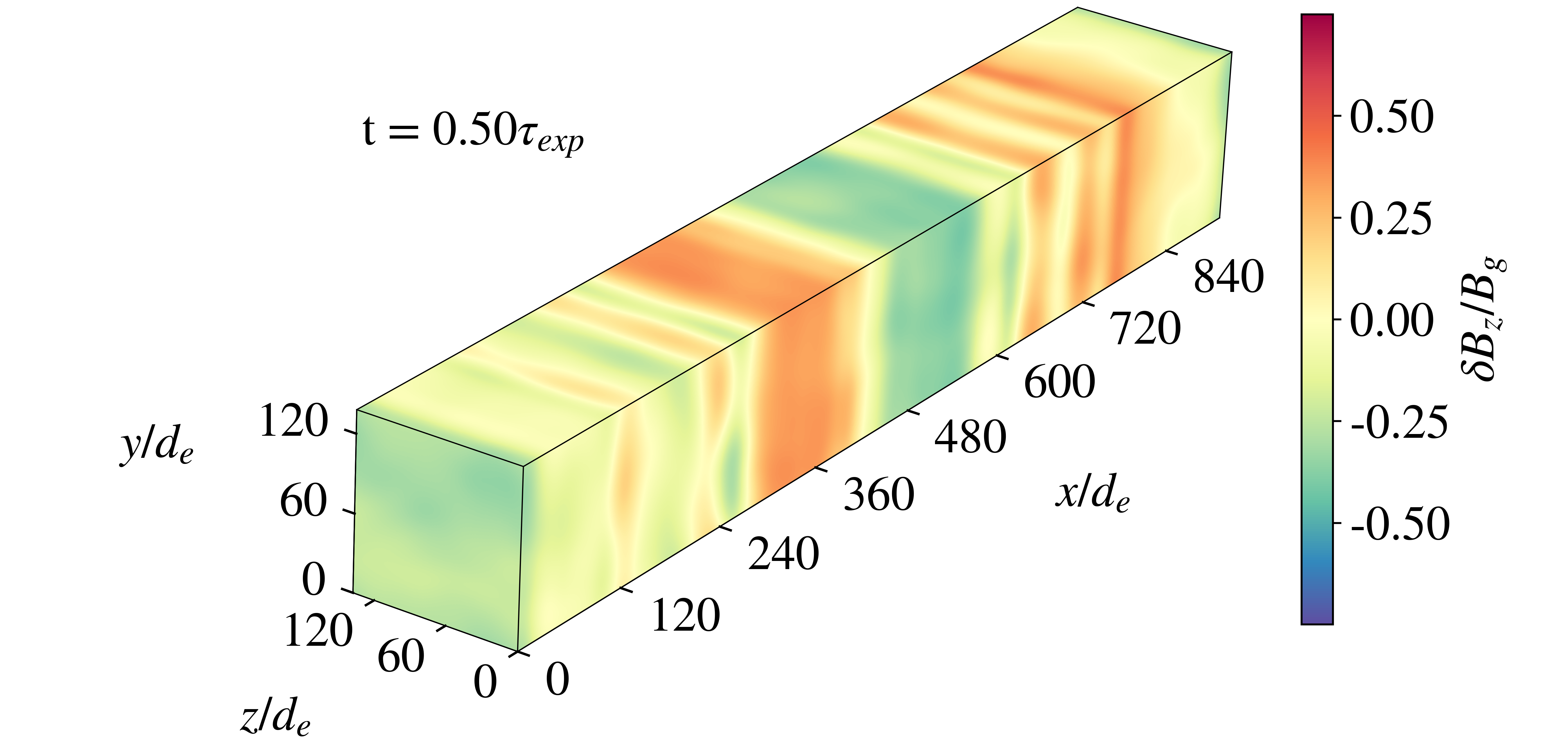}};
\node[anchor=north west] at (img.north west) {\small (b)};
\end{tikzpicture}
\end{minipage}

\par
\begin{minipage}[t]{0.5\linewidth}
\centering
\begin{tikzpicture}
\node[anchor=south west, inner sep=0] (img)
  {\includegraphics[width=\linewidth]{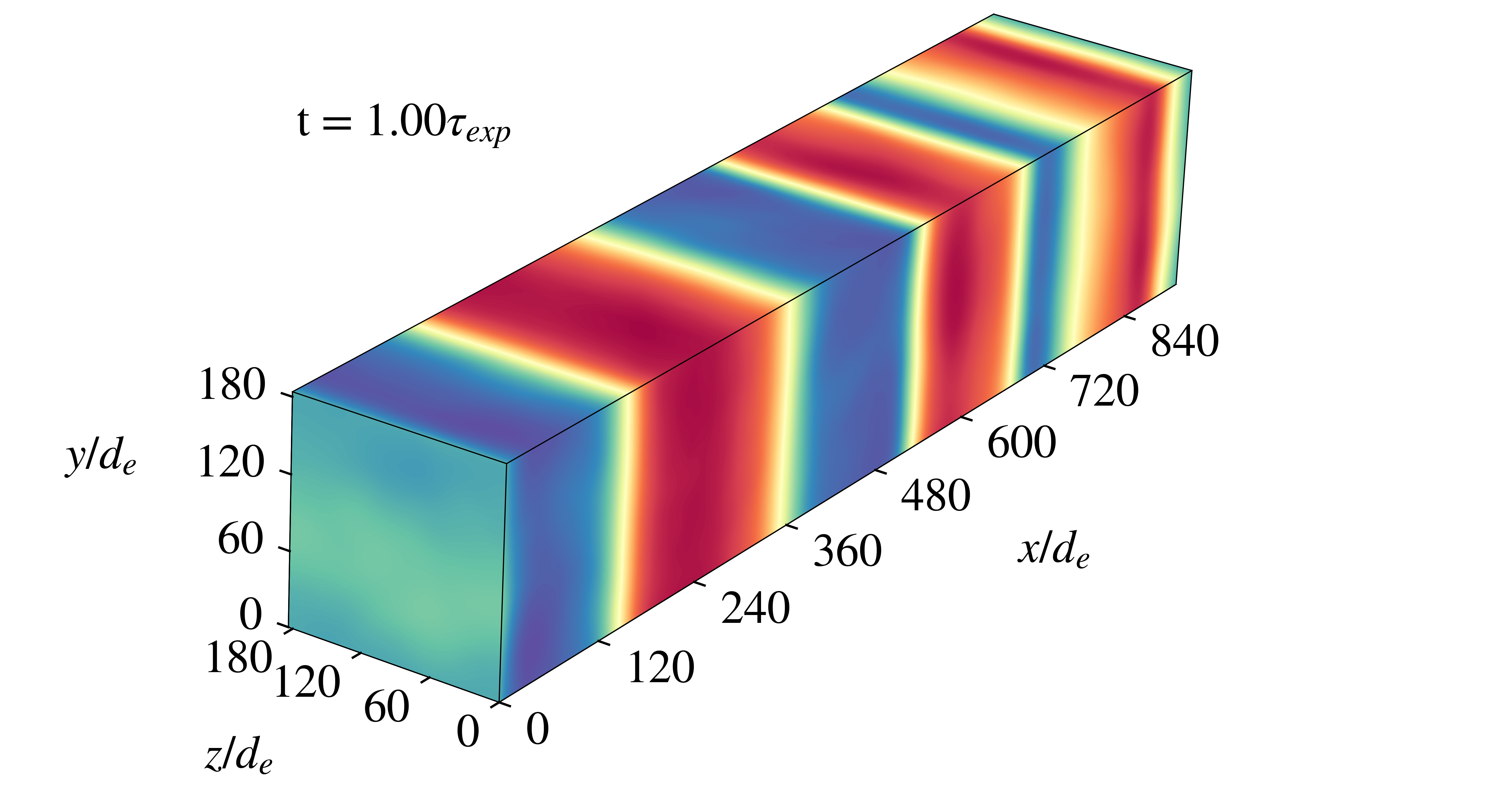}};
\node[anchor=north west] at (img.north west) {\small (c)};
\end{tikzpicture}
\end{minipage}%
\begin{minipage}[t]{0.5\linewidth}
\begin{tikzpicture}
\node[anchor=south west, inner sep=0] (img)
  {\includegraphics[
     height=0.53\linewidth
   ]{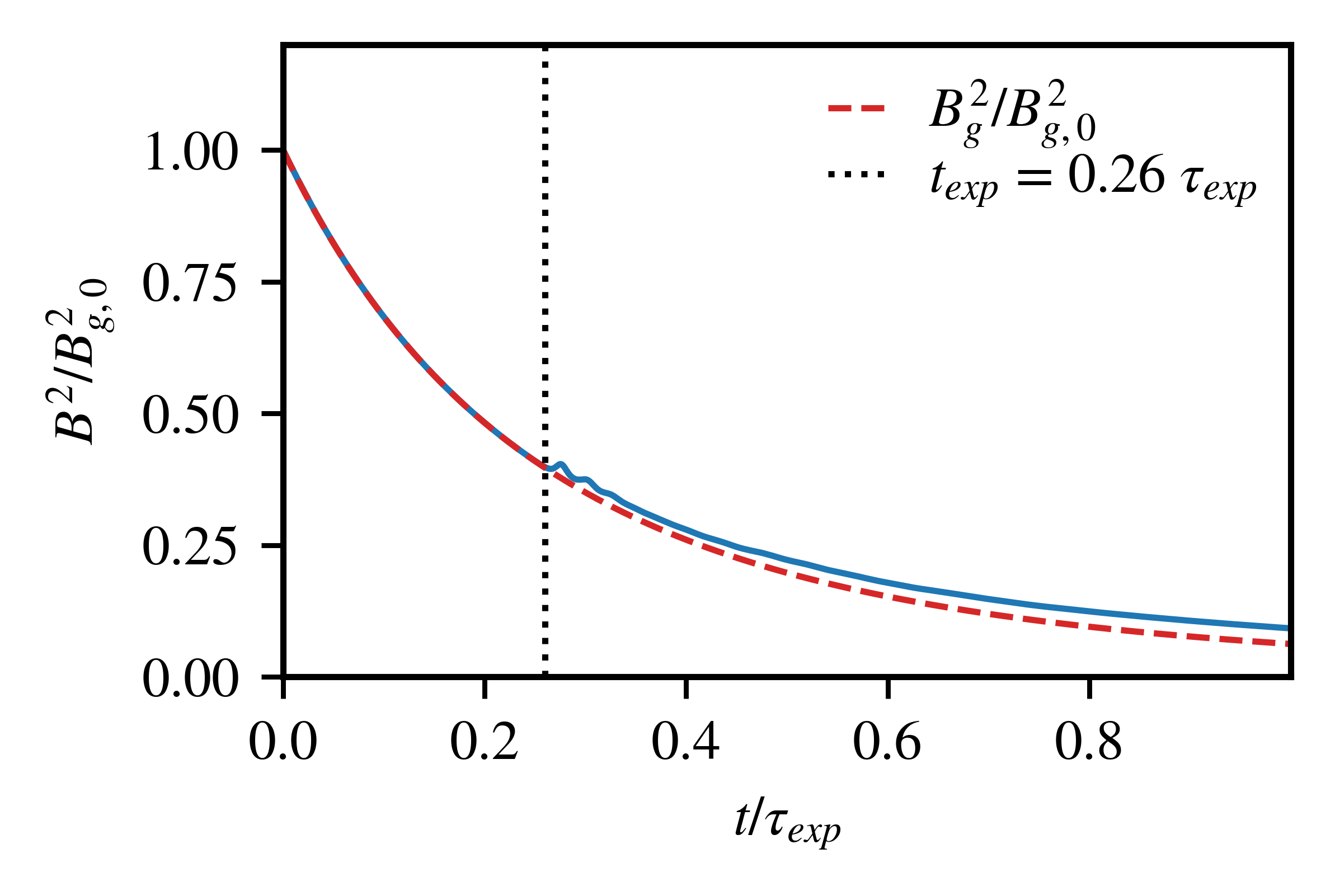}};
\node[anchor=north west] at (img.north west) {\small (d)};
\end{tikzpicture}
\end{minipage}
\caption{%
Three-dimensional visualization of the $z$-component of the magnetic-field fluctuations,
$\delta B_z/B_g$, shown at (a) the initial time, (b) half an expansion time, and (c) one
expansion time. Panel (d) shows the time evolution of the magnetic-field energy (solid blue) and the background guide-field energy (dashed red), normalized to the initial guide-field energy. The dotted black line marks the onset of firehose fluctuations.}
\label{fig:EB_combined}
\end{figure}

Fig.~\ref{fig:EB_combined} shows an example simulation conducted with the KEB. The simulation is initialized with a uniform background magnetic field $\vecB_{g,0}=(B_{g,0},0,0)$ such that the initial (ion) Alfv\'en speed $v_{\mathrm{A},0}\equiv B_{g,0}/\sqrt{4\pi m_i n_0}= 0.02c$. We initialize electrons and ions with a mass ratio $m_i/m_e=25$, uniformly distributed across the box and with velocities drawn from a Maxwellian with temperature $\theta_{0,i} = (m_e/m_i)\theta_{0,e} \equiv k_\mathrm{B}T_0/(m_ic^2)=1/2500$ such that the initial ion thermal speed is $\sim0.035c$, and the initial $\beta_i\equiv n_0 k_\mathrm{B}T/(B_{0,g}^2/(8\pi))=2$. The simulation domain has size $L_x\times L_y\times L_z = 184 \times 18.4 \times 18.4 d_i^3 $ at $t=0$, with $d_i \equiv c/\omega_{\mathrm{p},i}$ and $\omega_{\mathrm{p},i} =\sqrt{4 \pi n_0 e^2/m_i}$. Expansion is restricted to the perpendicular ($y$ and $z$) directions, and commences immediately upon starting the simulation. The expansion profile has rate $q_\mathrm{exp}$ such that one expansion time $\tau_\mathrm{exp}=1/q_\mathrm{exp}=91923.9\omega_{\mathrm{p},i}^{-1}$.

Fig.~\ref{fig:EB_combined}(a)--(c) shows the spatial distribution of magnetic-field fluctuations $\delta B_z$ at three stages during expansion. The magnetic field exhibits quasi-stationary parallel modes as the plasma becomes progressively unstable to the firehose instability. This is due to the temperature evolution, which changes as $T_\perp \propto B_g$ while $T_\parallel \sim \mathrm{const}$, creating pressure anisotropy. Furthermore, the parallel plasma-$\beta_\parallel \equiv n k_\mathrm{B} T_\parallel / (B_g^2/(8 \pi ))$ increases as $B_g$ decreases, naturally driving the plasma toward the firehose instability threshold. 
The simulation is run for 1 expasion time; the onset of the firehose instability can be seen in Fig.~\ref{fig:EB_combined}(d) at $t \sim 0.26 \tau_\mathrm{exp}$, where the total magnetic-field energy starts to diverge from the background magnetic-field energy (which evolves due to the box expansion) due to the emergence of the quasi-stationary parallel modes. 

\section{Leaky Box for PIC}
\label{sec:leak}

\subsection{Summary of the Leaky-Box algorithm}
Magnetized plasma turbulence can act as an efficient nonthermal particle accelerator. PIC simulations can be used to study turbulent acceleration of charged particles over their full history, including the injection phase of the acceleration. Moreover, PIC simulations can self-consistently capture the partitioning of energy between the thermal and nonthermal particles and the feedback between the turbulent cascade and cosmic rays. Out of computational convenience and in the desire of achieving maximum scale separation between the turbulence driving and dissipation scales, simulations routinely employ closed periodic boxes for fields and particles. Such systems typically suffer from the absence of a physically meaningful global energy sink. When energy is continuously supplied to maintain a turbulent state, the plasma internal energy essentially grows without bounds, whereas in more realistic settings sink terms provided by e.g.\ radiative cooling and/or particle escape balance the power input from the turbulent cascade. The lack of a global energy sink then manifests as a persistent ``drift'' of particle distributions to higher energies.  Moreover, at later stages of such simulations, nonthermal particles start piling up at the highest energies achievable for a given box size (i.e., at the Hillas limit).

A persistent steady state, however, can be achieved by introducing a diffusive escape mechanism for the particles. Here, we describe a recently developed method suitable for 
implementing diffusive particle escape in PIC simulations \cite{gorbunov2025prl}. The method does not interfere with periodic boundaries, used routinely in turbulence 
simulations, while introducing at the same time a form of escape that depends self-consistently on the properties of the turbulent flow. The method also incorporates the resupply of particles from an external medium, in order to maintain a steady particle number. The algorithm works as described below.

In order to track the diffusion at the particle level, we record the spatial displacement of each particle from its initial position in the box. 
Once the particle has diffused over some predefined distance $l_{\rm esc}$ it meets the condition for escape and is removed from the box. 
A new particle of the same mass and charge is then injected at the exact position of the removed particle, with initial velocity sampled from some predefined (typically Maxwellian) distribution. For simulations in the presence of a mean magnetic field $\vecB_0 =\{0,0,B_0\}$ with diffusive escape in directions perpendicular to $\vecB_0$, the algorithm can be summarized as:
\begin{enumerate}
    \item At the end of each timestep, compute the displacement vector $\Delta \bb{r}(t) = \{\Delta x(t), \Delta y(t)\} $ for each particle as 
    \begin{equation}
    \Delta \bb{r}^n = \Delta \bb{r}^{n-1} +  
    \frac{\bb{u}^n}{\gamma^n}\Delta t. 
    \end{equation}
    
    \item Check if
    \begin{equation}
    \max(|\Delta x|,|\Delta y|) \ge l_{\rm esc}. \label{eq:escape}    
    \end{equation}
    
    \item If condition~\eqref{eq:escape} is \emph{true}, then the particle velocity is resampled from a prescribed distribution function, and its displacement vector is reset to zero, $\Delta \bb{r}^n = \{0,0\}$. 
\end{enumerate}
In practice, we typically choose  $l_{\rm esc} = L/2$ for the diffusive escape distance, where $L$ is the linear size of the periodic box, although the algorithm can in principle be used with any reasonable choice of $l_{\rm esc}$. In periodic boxes, choices with $l_{\rm esc} \gtrsim L$ may suffer from spurious box-scale correlations. For accessing regimes with long particle confinement, it may thus be more appropriate to reduce the scale of the turbulent driver (relative to $L$) rather than increasing $l_{\rm esc}$ significantly beyond $L$. 
We also note that the escape condition, eq.~\eqref{eq:escape}, can be alternatively applied to the full displacement vector $\Delta \bb{r}  = \{\Delta x, \Delta y, \Delta z\}$. Our tests have shown that such a modification does not affect the simulation results, at least not in the case of large-amplitude turbulence with $\delta B/B_0\approx1$.

\subsection{Applications of the PIC Leaky Box}

\begin{figure*}[t]
    \centering \includegraphics[width=1\textwidth,trim={3cm 0 0 0},clip]{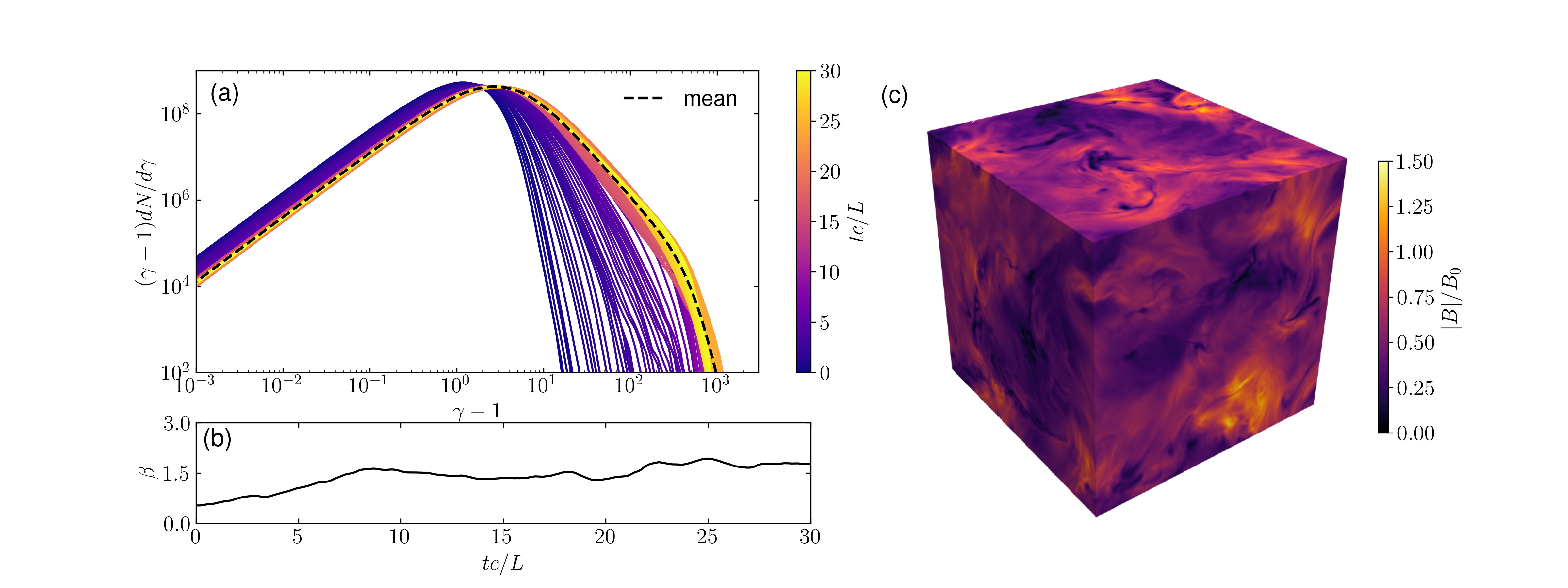}
    \caption{(a) Temporal evolution of the particle energy distribution function in a leaky-box turbulence simulation, where lighter colors indicate later simulation times. The dashed black line represents the time-averaged particle distribution during the steady state. (b) Temporal evolution of the volume-averaged plasma $\beta$. (c) Spatial distribution of the turbulent magnetic-field magnitude at a representative time during the steady state.}
    \label{fig:leakybox}
\end{figure*}

We demonstrate the application of the leaky-box scheme in Fig.~\ref{fig:leakybox}. The data used is from the simulations presented in ref.~\cite{gorbunov2025prl}.
We model a turbulent pair plasma with total number density $n_0$ threaded by an external guide field $B_0$, corresponding to a system with cold magnetization $\sigma_0\equiv B_0^2/(4\pi n_0m_ec^2)=1$. The temporal evolution of the particle energy distribution function is shown in Fig.~\ref{fig:leakybox}(a). After approximately $t =5 (L/c)$, a steady state is achieved, and the distribution functions do not significantly evolve further. The steady-state distribution features a prominent nonthermal high-energy tail which fluctuates around a mean, shown as a dashed black line in Fig.~\ref{fig:leakybox}(a). The system heats up until the ratio between the magnetic and thermal pressures settles to a state of $\beta\equiv n k_\mathrm{B} T/(B_0^2/(8\pi))\approx1.5$, as shown in Fig.~\ref{fig:leakybox}(b). The spatial distribution of the magnetic-field magnitude at a representative time during the steady state is shown in Fig.~\ref{fig:leakybox}(c).     

\section{Conclusion}
\label{sec:conclusion}

We have presented numerical methods to conduct fully kinetic simulations of plasmas where the standard PIC approach requires extensions to capture the correct physics. Standard PIC algorithms are modified to account for shearing, expansion, and escape, constructing new Maxwell solvers and particle pushers. For the particle-push step in particular, we have designed Boris-like algorithms that can be implemented as simple extensions of preexisting Boris pushers.

For simulations of sheared plasma, e.g.\ in accretion disks of compact objects where differential rotation imposes shearing, we presented the KSB-OA shearing-box algorithm. In this setup, a background shear mimicking differential (e.g.\ Keplerian) disk rotation is taken into account in the field equations and in the particle equations of motion. Gravitational and Coriolis forces are also included (but could also be excluded). The setup relies on shearing-periodic boundaries in the radial direction. In typical simulations, shearing an initial thermal plasma causes the development of the magnetorotational instability, which then (in 3D) results in sustained turbulence \cite{bacchini2022mri,bacchini2024prl,gorbunov2025apjl}.

For simulations of expanding plasmas, e.g.\ in the solar wind or in black-hole winds, we reviewed an expanding-box algorithm. The system's expansion is introduced as an extra force in the particle momentum equation and as frame-transformation factors in Maxwell's equations. An initial thermal plasma, when expanded, will typically experience the development of the firehose instability \cite{hellinger2019}, as we demonstrated.

In plasmas where particles are subjected to strong acceleration, e.g.\ in the surroundings of compact objects, a typical approach is to consider a PIC model of a local turbulent box. In such models, turbulence is usually driven by some external force, and the plasma continuously heats up due to the turbulent cascade. Accelerated particles are expected to leave the turbulent region and carry away energy, and we therefore introduced a leaky-box method in which particles are continuously removed and reinjected to mimic particle escape. This strategy allows us to obtain a true steady state in turbulence simulations \cite{gorbunov2025prl,groselj2026}. The method is also applicable to any closed-box PIC simulation with external energy injection, allowing the investigation of steady-state behavior in a number of standard PIC setups.

In future work, the algorithms discussed here could be further extended and, potentially, combined. For example, combining shearing and expansion/contraction would provide a better description of realistic collisionless accretion flows \cite{sironinarayan2015}. Adding the leaky-box strategy would additionally allow for the study of the associated steady state reached by particle energy distributions. This opens up many possible research pathways where more ingredients can be added to further extend PIC simulations and reach new level of realism.

\ack{We would like to thank Dmitri Uzdensky, Greg Werner, Vladimir Zhdankin, Lev Arzamasskiy, Sasha Philippov, Bart Ripperda, Lorenzo Sironi, Matt Kunz, Jim Stone, Elias Most, Andrea Mignone, Gianluigi Bodo, Maria Elena Innocenti, Alfredo Micera, sAgnieszka Wierzchucka, and Giuseppe Arr\`o for useful discussions throughout the development of this work.}

\funding{FB acknowledges support from the FED-tWIN programme (profile Prf-2020-004, project ``ENERGY''), issued by BELSPO, and from the FWO Junior Research Project G020224N granted by the Research Foundation -- Flanders (FWO).
PE is supported by the BELSPO BRAIN-be 2.0 Project B2/223/P1/PLATINUM. 
DG is supported by the Research Foundation--Flanders (FWO) Senior Postdoctoral Fellowship 12B1424N.}
The computational resources and services used in this work were provided by the VSC (Flemish Supercomputer Center), funded by the Research Foundation Flanders (FWO) and the Flemish Government, department EWI.


\data{Data presented in this work can be shared upon reasonable request.}



\begin{thebibliography}{}

\bibitem[{Bacchini} et~al., 2022]{bacchini2022mri}
{Bacchini}, F., Arzamasskiy, L., Zhdankin, V., Werner, G., Begelman, M., and Uzdensky, D. (2022).
\newblock {Fully Kinetic Shearing-box Simulations of Magnetorotational Turbulence in 2D and 3D. I. Pair Plasmas}.
\newblock {\em ApJ}, 938, 86.

\bibitem[Bacchini et~al., 2024]{bacchini2024prl}
Bacchini, F., Zhdankin, V., Gorbunov, E.~A., Werner, G.~R., Arzamasskiy, L., Begelman, M.~C., and Uzdensky, D.~A. (2024).
\newblock Collisionless magnetorotational turbulence in pair plasmas: Steady-state dynamics, particle acceleration, and radiative cooling.
\newblock {\em Phys. Rev. Lett.}, 133:045202.

\bibitem[Balbus and Hawley, 1991]{balbushawley1991}
Balbus, S.~A. and Hawley, J.~F. (1991).
\newblock {A Powerful Local Shear Instability in Weakly Magnetized Disks. I. Linear Analysis}.
\newblock {\em ApJ}, 376, 214.

\bibitem[Birdsall and Langdon, 1991]{Birdsall2005}
Birdsall, C.~K. and Langdon, A.~B. (1991).
\newblock {\em Plasma Physics via Computer Simulation}.
\newblock Taylor {\&} Francis Group, Boca Raton.

\bibitem[Boris, 1970]{boris1970}
Boris, J. (1970).
\newblock {Relativistic plasma simulation-optimization of a hybrid code}.
\newblock {\em Proceedings of the Fourth Conference on Numerical Simulations of Plasmas (Naval Research Laboratory, Washington DC, 1970), p. 3.}

\bibitem[Bott et~al., 2021]{bott2021}
Bott, A. F.~A., Arzamasskiy, L., Kunz, M.~W., Quataert, E., and Squire, J. (2021).
\newblock Adaptive critical balance and firehose instability in an expanding, turbulent, collisionless plasma.
\newblock {\em The Astrophysical Journal Letters}, 922, 2.

\bibitem[Dawson, 1983]{Dawson1983}
Dawson, J.~M. (1983).
\newblock {Particle Simulation of Plasmas}.
\newblock {\em Rev. Mod. Phys.}, 55(2):403.

\bibitem[Elena~Innocenti et~al., 2019]{innocenti2019method}
Elena~Innocenti, M., Tenerani, A., and Velli, M. (2019).
\newblock A semi-implicit particle-in-cell expanding box model code for fully kinetic simulations of the expanding solar wind plasma.
\newblock {\em The Astrophysical Journal}, 870(2):66.

\bibitem[Gorbunov et~al., 2025a]{gorbunov2025apjl}
Gorbunov, E.~A., Bacchini, F., Zhdankin, V., Werner, G., Begelman, M., , and Uzdensky, D. (2025a).
\newblock Leaking outside the box: Kinetic turbulence with cosmic-ray escape.
\newblock {\em ApJL}, 982, L28.

\bibitem[Gorbunov et~al., 2025b]{gorbunov2025prl}
Gorbunov, E.~A., Grošelj, D., and Bacchini, F. (2025b).
\newblock Leaking outside the box: Kinetic turbulence with cosmic-ray escape.
\newblock {\em Phys. Rev. Lett.}, 135:065201.

\bibitem[Gressel and Ziegler, 2007]{gresselziegler2007}
Gressel, O. and Ziegler, U. (2007).
\newblock {Shearingbox-implementation for the central-upwind, constraint-transport MHD-code NIRVANA}.
\newblock {\em CPC}, 176, 652.

\bibitem[{Groselj} et~al., 2026]{groselj2026}
{Groselj}, D., {Philippov}, A., {Beloborodov}, A.~M., and {Mushotzky}, R. (2026).
\newblock {High-energy Emission from Turbulent Electron-ion Coronae of Accreting Black Holes}.
\newblock {\em arXiv e-prints}, page arXiv:2601.00518.

\bibitem[Hellinger et~al., 2019]{hellinger2019}
Hellinger, P., Matteini, L., Landi, S., Franci, L., Verdini, A., and Papini, E. (2019).
\newblock Turbulence versus fire-hose instabilities: 3d hybrid expanding box simulations.
\newblock {\em The Astrophysical Journal}, 883(2):178.

\bibitem[Hoshino, 2013]{hoshino2013}
Hoshino, M. (2013).
\newblock {Particle acceleration during magnetorotational instability in a collisionless accretion disk}.
\newblock {\em ApJ}, 773, 118.

\bibitem[Hoshino, 2015]{hoshino2015}
Hoshino, M. (2015).
\newblock {Angular Momentum Transport and Particle Acceleration During Magnetorotational Instability in a Kinetic Accretion Disk}.
\newblock {\em PRL}, 114, 061101.

\bibitem[Inchingolo et~al., 2018]{inchingolo2018}
Inchingolo, G., Grismayer, T., Loureiro, N.~F., Fonseca, R.~A., and Silva, L.~O. (2018).
\newblock {Fully Kinetic Large-scale Simulations of the Collisionless Magnetorotational Instability}.
\newblock {\em ApJ}, 859, 149.

\bibitem[Innocenti et~al., 2019]{innocenti2019}
Innocenti, M.~E., Tenerani, A., Boella, E., and Velli, M. (2019).
\newblock Onset and evolution of the oblique, resonant electron firehose instability in the expanding solar wind plasma.
\newblock {\em The Astrophysical Journal}, 883(2):146.

\bibitem[Kunz et~al., 2014]{kunz2014code}
Kunz, M.~W., Stone, J.~M., and Bai, X. (2014).
\newblock {Pegasus: A new hybrid-kinetic particle-in-cell code for astrophysical plasma dynamics}.
\newblock {\em JCP}, 259, 154.

\bibitem[Kunz et~al., 2016]{kunz2016}
Kunz, M.~W., Stone, J.~M., and Quataert, E. (2016).
\newblock {Magnetorotational Turbulence and Dynamo in a Collisionless Plasma}.
\newblock {\em PRL}, 117, 235101.

\bibitem[Lapenta, 2012]{lapenta2012}
Lapenta, G. (2012).
\newblock Particle simulations of space weather.
\newblock {\em Journal of Computational Physics}, 231(3):795--821.
\newblock Special Issue: Computational Plasma Physics.

\bibitem[Micera et~al., 2021]{micera2021}
Micera, A., Zhukov, A.~N., López, R.~A., Boella, E., Tenerani, A., Velli, M., Lapenta, G., and Innocenti, M.~E. (2021).
\newblock On the role of solar wind expansion as a source of whistler waves: Scattering of suprathermal electrons and heat flux regulation in the inner heliosphere.
\newblock {\em The Astrophysical Journal}, 919(1):42.

\bibitem[Nishikawa et~al., 2021]{nishikawa2021}
Nishikawa, K., Duţan, I., Köhn, C., and Mizuno, Y. (2021).
\newblock {PIC methods in astrophysics: simulations of relativistic jets and kinetic physics in astrophysical systems}.
\newblock {\em Living Rev Comput Astrophys}, 7, 1.

\bibitem[Ripperda et~al., 2018]{ripperda2018a}
Ripperda, B., Bacchini, F., Teunissen, J., Xia, C., Porth, O., Sironi, L., Lapenta, G., and Keppens, R. (2018).
\newblock A comprehensive comparison of relativistic particle integrators.
\newblock {\em ApJS}, 235, 21.

\bibitem[Riquelme et~al., 2012]{riquelme2012}
Riquelme, M.~A., Quataert, E., Sharma, P., and Spitkovsky, A. (2012).
\newblock {Local two-dimensional particle-in-cell simulations of the collisionless magnetorotational instability}.
\newblock {\em ApJ}, 755, 50.

\bibitem[Sandoval et~al., 2024]{sandoval2024}
Sandoval, A., Riquelme, M., Spitkovsky, A., and Bacchini, F. (2024).
\newblock {Particle-in-cell simulations of the magnetorotational instability in stratified shearing boxes}.
\newblock {\em Monthly Notices of the Royal Astronomical Society}, 530(2):1866--1884.

\bibitem[Sironi and Narayan, 2015]{sironinarayan2015}
Sironi, L. and Narayan, R. (2015).
\newblock Electron heating by the ion cyclotron instability in collisionless accretion flows. i. compression-driven instabilities and the electron heating mechanism.
\newblock {\em The Astrophysical Journal}, 800, 2.

\bibitem[Squire et~al., 2020]{squire2020}
Squire, J., Chandran, B. D.~G., and Meyrand, R. (2020).
\newblock In-situ switchback formation in the expanding solar wind.
\newblock {\em The Astrophysical Journal Letters}, 891(1):L2.

\bibitem[Stone and Gardiner, 2010]{stonegardiner2010}
Stone, J.~M. and Gardiner, T.~A. (2010).
\newblock {Implementation of the shearing box approximation in ATHENA}.
\newblock {\em ApJS}, 189, 142.

\bibitem[Tran et~al., 2023]{tran2023}
Tran, A., Sironi, L., Ley, F., Zweibel, E.~G., and Riquelme, M.~A. (2023).
\newblock Electron reacceleration via ion cyclotron waves in the intracluster medium.
\newblock {\em The Astrophysical Journal}, 948, 2.

\bibitem[Villasenor and Buneman, 1992]{villasenorbuneman1992}
Villasenor, J. and Buneman, O. (1992).
\newblock {Rigorous charge conservation for local electromagnetic field solvers}.
\newblock {\em CPC}, 69, 306.

\end{thebibliography}

\end{document}